\documentclass[reprint,amsmath,amssymb,aps,superscriptaddress,prstab]{revtex4-1}
\usepackage{graphicx}% Include figure files
\usepackage{dcolumn}% Align table columns on decimal point
\usepackage{bm}% bold math
\usepackage{siunitx}
\usepackage{color}
\usepackage{amsmath}
\usepackage{comment}
\usepackage{url} 
\usepackage[utf8x]{inputenc}

\usepackage[colorlinks, citecolor=blue, urlcolor=blue, bookmarks=false, linkcolor=blue, hypertexnames=true]{hyperref}

\def\ss#1{\textcolor{black}{#1}}

\begin{document}
\title{Lattice dynamics and magnetic exchange interactions in GeCo$_2$O$_4$, a spinel with S = 1/2 pyrochlore lattice}

\author{Prativa Pramanik\textsuperscript{*}}
\affiliation{Department of Physics, Indian Institute of Technology Guwahati, Assam 781039, India}

\author{Sobhit Singh\textsuperscript{*,\textdagger}}
\affiliation{Department of Physics and Astronomy, Rutgers University, Piscataway, New Jersey 08854-8019, USA}
%\email{sobhit.singh@rutgers.edu}

\author{Mouli Roy Chowdhury}
\affiliation{Department of Physics, Indian Institute of Technology Guwahati, Assam 781039, India}

\author{Sayandeep Ghosh}
\affiliation{Department of Physics, Indian Institute of Technology Guwahati, Assam 781039, India}

\author{Vasant Sathe}
\affiliation{UGC DAE Consortium for Scientific Research, Indore- 452 001, India}

\author{Karin M. Rabe}
\affiliation{Department of Physics and Astronomy, Rutgers University, Piscataway, New Jersey 08854-8019, USA}

\author{David Vanderbilt}
\affiliation{Department of Physics and Astronomy, Rutgers University, Piscataway, New Jersey 08854-8019, USA}

\author{Mohindar S. Seehra}
\affiliation{Department of Physics \& Astronomy, West Virginia University, Morgantown, WV 26506, USA}

\author{Subhash Thota\textsuperscript{\ddag}}
\affiliation{Department of Physics, Indian Institute of Technology Guwahati, Assam 781039, India}
%\email{subhasht@iitg.ac.in}

\begin{abstract}
\ss{GeCo\textsubscript{2}O\textsubscript{4}} is a unique system in the family of cobalt spinels ACo\textsubscript{2}O\textsubscript{4} (A= Sn, Ti, Ru, Mn, Al, Zn, Fe, etc.) in which magnetic Co ions stabilize on the  pyrochlore lattice exhibiting a large degree of orbital frustration. Due to the complexity of the low-temperature antiferromagnetic (AFM) ordering and long-range magnetic exchange interactions, the lattice dynamics and magnetic structure of GeCo\textsubscript{2}O\textsubscript{4} spinel has remained puzzling. To address this issue, here we present theoretical and experimental investigations of the highly frustrated magnetic structure, and the infrared (IR) and Raman-active phonon modes in the spinel GeCo\textsubscript{2}O\textsubscript{4}, which exhibits an AFM ordering below the N{\'e}el temperature \emph{T}\textsubscript{N} $\sim$21\,K 
\ss{and an associated cubic ($Fd{\bar 3}m$) to tetragonal ($I4_{1}/amd$) structural phase transition whose location at \emph{T}\textsubscript{N} vs. at a lower \emph{T}\textsubscript{S} $\sim$16\,K is controversial. }
%, followed by a cubic ($Fd{\bar 3}m$) to tetragonal ($I4_{1}/amd$) structural phase transition at \emph{T}\textsubscript{S} $\sim$16\,K. 
%
Our density-functional theory (DFT+$U$) calculations reveal that one needs to consider magnetic-exchange interactions up to the third nearest neighbors to get an accurate description of the low-temperature AFM order in  GeCo\textsubscript{2}O\textsubscript{4}. At room temperature three distinct IR-active modes (\emph{T}\textsubscript{1u}) are observed at frequencies 680, 413, and 325 cm$^{-1}$ along with four Raman-active modes \emph{A}\textsubscript{1g}, \emph{T}\textsubscript{2g}(1), \emph{T}\textsubscript{2g}(2), and  \emph{E}\textsubscript{g}  at frequencies 760, 647, 550, and 308\,cm$^{-1}$, respectively, which match reasonably well with our DFT+$U$ calculated values. All the IR-active and Raman-active phonon modes exhibit signatures of moderate spin-phonon coupling. The temperature dependence of various parameters, such as the shift, width, and intensity, of the Raman-active modes, is also discussed. 
Noticeable changes around \ss{\emph{T}\textsubscript{N}$\sim$ 21\,K and \emph{T}\textsubscript{S} $\sim$ 16\,K} are observed in the Raman line parameters of the \emph{E}\textsubscript{g} and \emph{T}\textsubscript{2g}(1) modes, which are associated with the modulation of the Co-O bonds in CoO\textsubscript{6} octahedra during the excitations of these modes. 
\end{abstract}

\maketitle

\section{Introduction}
\vspace{-0.3cm}
The diversity in the properties and applications of spinels with the general formula of AB\textsubscript{2}O\textsubscript{4} arises from the variety of cations, magnetic or nonmagnetic, that can be substituted at the tetrahedral A-sites and octahedral B-sites of the spinel structure~\cite{morrish2001physical,Rabe_review2010, thanh2012magnetic,ChakhalianRMP2014, seehra2017magnetic, thota2017nature, SinghJAP2017, Pramanik_2017, harris2009recent}. 
%Recent studies on a special subclass of spinels \ss{{\it viz.,} having nonmagnetic cations such as Zn\textsuperscript{2+}, Mg\textsuperscript{2+}, Ge\textsuperscript{4+}, etc. at the A-sites and magnetic cations at the B-sites, reveal} 
\ss{Recent studies on a subclass of spinels having nonmagnetic cations such as Zn\textsuperscript{2+}, Mg\textsuperscript{2+}, and Ge\textsuperscript{4+} at the A-sites and magnetic cations at the B-sites reveal} 
intriguing magnetic and structural properties at low temperatures. As first pointed out by Anderson~\cite{anderson1956ordering}, these spinels have inherent magnetic frustration, making the long-range magnetic order, if at all present, highly dependent on various other factors~\cite{ChakhalianRMP2014,seehra2017magnetic,thota2017nature, harris2009recent,LiuNL2019}. %
Examples of such spinels are ZnFe\textsubscript{2}O\textsubscript{4}~\cite{schiessl1996magnetic}, defect spinel MgMnO\textsubscript{3}~\cite{seehra2011magnetic}, and GeCo\textsubscript{2}O\textsubscript{4}~\cite{GhoshPRB2018, pramanik2019magnetic}. The latter is the subject of this paper. It is noteworthy that GeCo\textsubscript{2}O\textsubscript{4}, hereafter listed as GCO for brevity, has been substantially investigated in connection with its use as an anode material for Li-ion batteries~\cite{ge2012co,jin2015ultrathin,subramanian2017facile,yuvaraj2017electrochemical}. Moreover, the nanostructures of GCO have found applications in the renewable energy sectors such as fuel-cells, electrochemical sensors, and supercapacitors~\cite{jin2015ultrathin,yuvaraj2017electrochemical}.

The magnetic properties of GCO have been under intense investigation in recent years because of the distinct magnetoelectric features linked to the noncollinear spin arrangement and distorted cubic structure. Based in part on several previous electron-spin resonance, magnetic, and neutron diffraction studies in GCO~\cite{Okubo2017,Yamasaki_2012, diaz2006magnetic, matsuda2011magnetic, horibe2006spontaneous, lashley2008specific,barton2014structural,fabreges2017field,tomiyasu2011molecular, GhoshPRB2018}, Pramanik {\it et al.}~\cite{pramanik2019magnetic} \ss{recently} presented results on the magnetic ground state, magnetic-field-induced transitions, and optical bandgap of GCO. 
Summarizing these results, \ss{it was shown that GCO contains a pyrochlore lattice of Co\textsuperscript{2+} spin moments which have effective spin \emph{S} = 1/2 (instead of \emph{S} = 3/2 as expected from the Hund's rules) due to the effects of the spin-orbit coupling and Jahn-Teller distortion. The magnetic ordering consists} 
%
%contains a  pyrochlore lattice of Co\textsuperscript{2+} spin moments with effective spin \emph{S} = 1/2, consisting 
of alternate planes of kagom\'e (KGM) and triangular (TRI) spins lying perpendicular to the {[}111{]} direction. 
The dominant in-plane exchange constant between the Co\textsuperscript{2+} spins is ferromagnetic (FM). However, the spins in the neighboring planes are ordered antiferromagnetically (AFM) with $\textbf{q}$ = ($\frac{1}{2},\frac{1}{2},\frac{1}{2}$) to yield an overall AFM order in the absence of any external magnetic field below the N\'eel temperature \emph{T}\textsubscript{N} = 20.4\,K~\cite{pramanik2019magnetic}. 
Due to such a peculiar magnetic behavior, especially owing to the frustrated AFM ordering  with $\textbf{q}$ = ($\frac{1}{2},\frac{1}{2},\frac{1}{2}$), various exotic competing magnetic phases such as classical and quantum spin liquid phases, recently reported in (111)-oriented quasi-two-dimensional spinels through a geometric lattice design approach, can be realized in GCO at low temperatures~\cite{LiuNL2019, liu2021proximate,ChakhalianAPL2020}.

Several studies reported a cubic ($Fd{\bar 3}m$) to tetragonal ($I4_{1}/amd$) distortion of the lattice accompanying \emph{T}\textsubscript{N} ~\cite{lashley2008specific,hoshi2007magnetic,watanabe2008jahn}, although high resolution x-ray diffraction studies by Barton \emph{et al.}~\cite{barton2014structural} revealed that the tetragonal distortion
\ss{of $\sim$0.1\% in the lattice parameters} 
occurs at \emph{T\textsubscript{S}} = 16\,K, a few degrees below \emph{T}\textsubscript{N}, \ss{along with modulation of the
Co-O bonds in the CoO$_6$ octahedra. However, it likely has} 
a nonmagnetic origin since no anomalies occur in the heat capacity \ss{and magnetic susceptibility} data near \emph{T\textsubscript{S}}~\cite{barton2014structural}. 
Also, the degree of tetragonality progressively increases with decreasing temperature~\cite{barton2014structural}. This cubic-to-tetragonal structural phase transition was attributed to local Jahn-Teller  effects~\cite{barton2014structural}, which lift the degeneracy of the $t_{2g}$ states by minimizing the energy of the $d_{xz}$ and $d_{yz}$ Co-3$d$ sub-orbitals~\cite{Ghosh_2021}. 
The closeness between the magnetic and structural transition temperatures reveals the existence of competing spin-orbit coupling and Jahn-Teller effects in GCO~\cite{barton2014structural,hoshi2007magnetic,watanabe2008jahn, Ghosh_2021}. Currently, there exists a fair amount of debate regarding the fact that $T_S$ is below the $T_N$, which is uncommon when compared to other spinels that exhibit magnetostructural quantum phase transitions~\cite{bordacs2009magnetic, thota2014ac, kim2012giant, kim2011pressure, guillou2011magnetic, suchomel2012spin, thota2017neutron,nayak2015magnetic,nayak2016low,nayak2016reentrant,pramanik2020neutron,thota2015nature,thota2013co}.
 
A systematic investigation of the temperature-dependent lattice dynamics is required to pin down the nature of transitions occurring near $T_S$ and $T_N$ in GCO. The only previously reported Raman studies in GCO are those of Koringstein \emph{et al.}~\cite{koningstein1972light}, which reported the observation of three Raman-active modes (\emph{A}\textsubscript{1g}, \emph{T}\textsubscript{2g}(1), and \emph{E}\textsubscript{g}) in GCO. However, these studies were done at only two temperatures, 200 and 400\,K, which are much higher than the $T_S$ and $T_N$. Also, the only yet reported infrared (IR) study in GCO was performed by Preudhomme and Tarte~\cite{preudhomme1972infrared} at 300\,K, which reported the observation of four IR-active modes (\emph{T}\textsubscript{1u}). 

In this work, we perform detailed temperature-dependent Raman measurements covering the temperature range \ss{of} 5 to 300 K with a focus on the changes occurring in the Raman-active modes as the temperature is lowered through  \emph{\emph{T}\textsubscript{N}} and \emph{T\textsubscript{S}}. Notably, our low-temperature Raman measurements confirm that the structural phase transition in GCO follows the magnetic phase transition, as first reported by Barton \emph{et al.}~\cite{barton2014structural} using x-ray diffraction measurements. We observe noticeable changes in the line parameters of the \emph{E}\textsubscript{g} and \emph{T}\textsubscript{2g} modes, which are associated with the modulation of the Co-O bonds in CoO\textsubscript{6} octahedra, near the \emph{T}\textsubscript{N} and \emph{T}\textsubscript{S}. We further report the observation of three (out of four) symmetry-allowed IR-active \emph{T}\textsubscript{1u} modes along with two satellite modes likely appearing due to the local symmetry breaking. In addition, computational studies of the lattice modes using density-functional theory (DFT+$U$) calculations are presented, revealing the presence of moderate spin-phonon coupling in GCO. A systematic analysis of the Heisenberg spin Hamiltonian suggests that the magnetic-exchange interactions up to the third nearest neighbors are required to accurately describe the low-temperature AFM ordering in GCO. Besides, we also briefly comment on the problems encountered in the DFT+$U$ calculations involving the orbital occupation of Co-$3d$ orbitals located at the magnetically frustrated sites in GCO. 

This paper is organized as follows.
\ss{In Sec.~\ref{sec:methods}, experimental and computational details of this study are
presented.}
Section~\ref{sec:resultsdiscussion} contains all the results and discussions in the following order: first, we discuss the crystal structure and the magnetic-exchange interactions in GCO, and then, we present our theoretical and experimental investigations on the lattice dynamics of GCO. This is  followed by conclusions in Sec.~\ref{sec:conclusions}.

\section{Methods}
\label{sec:methods}

\subsection{Experimental details}
\label{sec:exp_details}
\vspace{-0.2cm}

A well-grounded mixed powder of high purity GeO\textsubscript{2}
(Sigma-Aldrich, 99.99\%) and Co\textsubscript{3}O\textsubscript{4}
(Sigma-Aldrich, 99.99\%) in stoichiometric amounts was pressed \ss{into} a cylindrical disc at 50
kg/cm\textsuperscript{2} by hydraulic press and followed by the sintering
process to yield the desired compound. The details of the sample synthesis procedures are described in a previous publication~\cite{pramanik2019magnetic}. 
The single phase of the synthesized sample was confirmed by x-ray diffraction measurements using a high-resolution XPERT-PRO diffractometer
(Co-K\textsubscript{$\alpha$} radiation with $\lambda$ = 1.78901 \AA). 
The temperature-dependent vibrational Raman-scattering spectra of
GCO were recorded with a commercial Labram-HR800 micro-Raman spectrometer, in the temperature range of 5\,K to 300\,K, using a He--Ne laser of wavelength 514\,nm. 
For frequency calibration the silicon mode at 520 cm$^{-1}$ was used. 
All the Raman spectra were recorded in the anti-Stokes region. 
For the low-temperature measurements, the sample was first mounted on a cold stage
setup (THMS600 stage from Linkam UK) equipped with a temperature
controller capable of maintaining a steady temperature. 
The sample was cooled by liquid helium and the temperature controller was able to hold
the temperature fluctuations within a range of $\pm$1\,K. 
\ss{The experimental uncertainty in the Raman peak positions, as determined using Lorentzian oscillator fits, was less than 0.1 cm$^{-1}$.}
The room temperature IR spectrum was recorded using a Perkin-Elmer
Spectrum-Two system with the standard spectral resolution of 0.5\,cm$^{-1}$. \ss{The IR-active mode frequencies were determined by Lorentzian oscillator fits of the transmittance data.} 

%\vspace{-0.5cm}
\subsection{Computational details}
\label{sec:comp_details}
\vspace{-0.2cm}

In order to better understand the nature of the magnetic-exchange interactions and  Raman and IR-active phonon modes in GCO, we carried out DFT+$U$ based first-principles calculations using the Projector Augmented Wave (PAW) method as implemented in the VASP software~\cite{Kresse96a, Kresse96b, KressePAW}. 
The PAW pseudopotentials considered the following valence configurations: 
Ge 4s\textsuperscript{2}4p\textsuperscript{2}, Co 3d\textsuperscript{8}4s\textsuperscript{1}, and O 2s\textsuperscript{2}2p\textsuperscript{4}. 
A kinetic energy cutoff of 650\,eV was set for the plane waves. The reciprocal space was sampled using a Monkhorst-pack k-mesh~\cite{MP1976} of size 8$\times$8$\times$8. The energy convergence criterion for the self-consistent DFT+$U$  calculations was set to 10$^{-7}$\,eV, and the force convergence criterion for relaxation calculations was set to 10$^{-3}$\,eV/\AA. All DFT+$U$ calculations were performed for collinear magnetic configurations without considering spin-orbit coupling effects. 
\ss{{\sc PyProcar} software~\cite{pyprocar} was used to plot the density of states, shown in Supplemental Materials (SM)~\cite{SM}.} 
We used the {\sc Phonopy} package to study the lattice dynamics~\cite{phonopy}. 
Supercells of size 2$\times$2$\times$2 were employed to calculate the phonon frequencies and phonon eigenvectors within the finite-displacement approach.
The exchange-correlation functional was computed using the generalized-gradient approximation (GGA) as parameterized by Perdew-Burke-Ernzerhof (PBE) as well as the PBE revised for solids (PBEsol)~\cite{PBE, PBEsol}. 
We find that the PBEsol yields lattice parameters and phonon frequencies that are in better agreement with the experimental data as compared to the PBE predictions. 

\ss{The onsite-Coulomb interaction effects for Co-$3d$ electrons were treated at the mean-field level using the rotationally invariant DFT+$U$ method introduced by Liechtenstein {\it et al.}~\cite{Liechtenstein1995}. 
We set $U$ = 4.0\,eV and $J$ = 1.0\,eV~\cite{GhoshPRB2018}. We find that this set of values appropriately describes the lattice parameters, magnetic structures, and vibrational properties of GCO. No tuning of the ($U, J$) parameters was performed to match the calculated phonon frequencies with the experimental data. Besides, it has been reported that an effective $U$\textsubscript{eff}\,\textsuperscript{Co} (= $U-J$) in the range of 2 – 3 eV provides a reasonable prediction of the electronic structure and optical properties of GCO~\cite{GhoshPRB2018}.}

We often noticed an anomalous variation in the occupation of the Co-3$d$ orbitals in some of our DFT+$U$ calculations due to the presence of strong magnetic frustration effects leading to a metastability problem in this system~\cite{MeredigPRB2010, AllenWatson2014}. To ensure the correct and consistent occupation of the Co-3$d$ orbitals, we utilized the occupation matrix control methodology developed by Allen and Watson~\cite{AllenWatson2014} in our reported  DFT+$U$ calculations. 
We optimized the structural primitive cell in the FM order since the FM order preserves the cubic symmetry of the paramagnetic phase. The PBE+$U$ and PBEsol+$U$ optimized lattice parameters are 8.434 and 8.322\,\AA, respectively. We observed that the PBEsol+$U$ optimized lattice parameters are in excellent agreement with the reported experimental data (8.3191\,\AA)~\cite{pramanik2019magnetic, barton2014structural}. Further, the PBEsol+$U$ optimized Co$-$O and Ge$-$O bond lengths are 2.1\,\AA~and 1.8\,\AA, respectively, which agree very well with the reported experimental data (2.1\,\AA~and 1.8\,\AA)~\cite{pramanik2019magnetic,Yamasaki_2012, barton2014structural}.

\section{Results and Discussion}
\label{sec:resultsdiscussion}

\subsection{Crystal structure and magnetic structure of G\MakeLowercase{e}C\MakeLowercase{o}$_2$O$_4$}

\begin{figure}[h!]
    \centering
    \includegraphics[width=0.95\columnwidth]{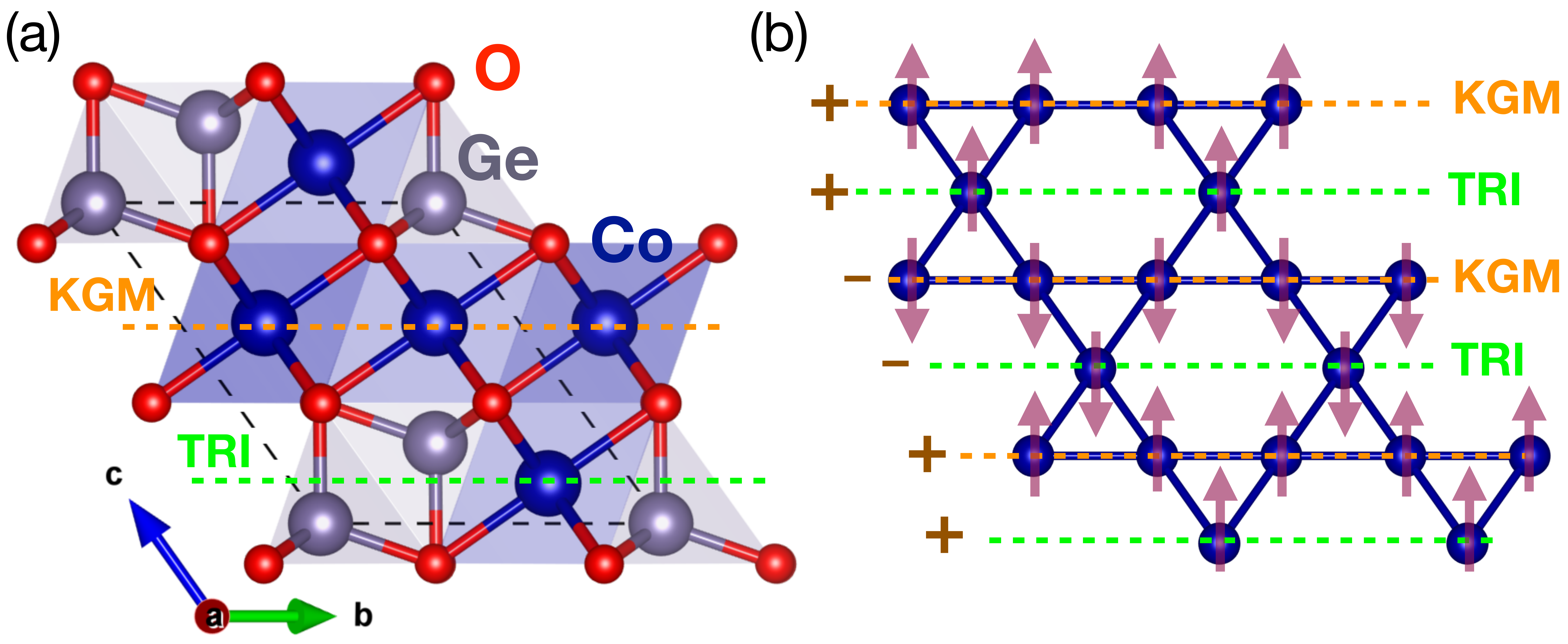}
    \caption{(a) Crystal structure of GCO. Dashed black lines mark the boundaries of the structural primitive cell. The primitive cell consists of two GeO\textsubscript{4} tetrahedra and one Co\textsubscript{4} tetrahedral unit.
    Here Co, Ge, and O ions are presented by the blue, gray, and red color, respectively.
    (b) Schematic representation of $\rm{\bf{q}}$ = ($\frac{1}{2}, \frac{1}{2}, \frac{1}{2}$) AFM ordering in GCO. Magnetic moments at Co sublattice are shown using  arrows in a collinear setting, {\it i.e.}, majority or up ($+$) and minority or down ($-$) spin states are denoted using up and down arrows, respectively.
    Alternating kagom\'e (KGM) and triangular (TRI) planes are highlighted using dashed horizontal lines. 
    A $T_{+}\,K_{+}\,T_{-}\,K_{-}\,T_{+}\,K_{+}\,T_{-}\,K_{-}\cdots$ type AFM spin configuration of TRI ($T_{\pm}$) and KGM ($K_{\pm}$) layers can be noticed along the [111] direction. 
    }
    \label{fig:fig1struct}
\end{figure}

GeCo$_2$O$_4$, [(Ge$^{4+})_{A}$[Co$^{2+}_{2}]_{B}$O$_{4}$],  crystallizes in a normal cubic spinel structure at room temperature (space group $Fd\bar{3}m$). 
The oxygen anions are located at the 32$e$ Wyckoff positions forming a close-packed face-centered cubic arrangement, whereas Ge and Co cations occupy the 8$a$-tetrahedral and 16$d$-octahedral interstitial positions, respectively. 
Therefore, the crystal structure consists of the corner sharing CoO$_6$ octahedra and GeO$_4$ tetrahedra, as shown in Fig.~\ref{fig:fig1struct}(a). 
The structural primitive cell contains 2 formula units of GCO. There are 4 magnetic Co atoms in the primitive cell forming a regular Co-Co tetrahedron, where each Co is located at the center of an oxygen octahedron at the 16$d$ sites. The corner-sharing oxygen octahedra form a pyrochlore lattice containing alternating planes of the KGM and TRI layers of Co atoms stacked along the [111] direction of the bulk unit cell. 

There are 3 Co atoms in the KGM plane and 1 Co atom in the TRI plane, as shown in Fig.~\ref{fig:fig1struct}(a).  
Within each KGM and TRI planes, Co spins order ferromagnetically. However, the overall low-temperature magnetic structure of GCO is much complex involving an antiferromagnetic ordering of wave vector $\rm{\bf{q}}$ = ($\frac{1}{2}, \frac{1}{2}, \frac{1}{2}$). 
In this AFM order, Co spins in a pair of TRI and KGM layers ($i.e.$, within a structural primitive cell) order ferromagnetically, whereas the same order antiferromagnetically in the neighboring structural primitive cell, thus 
resulting in a TRI-KGM  layer spin configuration of $T_{+}\,K_{+}\,T_{-}\,K_{-}\,T_{+}\,K_{+}\,T_{-}\,K_{-}\cdots$ along the [111] direction, as shown in Fig.~\ref{fig:fig1struct}(b). 
Here, $T_{+}$ ($T_{-}$) and $K_{+}$ ($K_{-}$) denote the spin up (down) configurations of the TRI and KGM layers, respectively.

%%%%%%%%%%%%%%%%%%%%%%%%%%%%%
%%%%%%%%%%%%%%%%%%%%%%%%%%%%%

\begin{figure}[htb!]
    \centering
    \includegraphics[width=0.95\columnwidth]{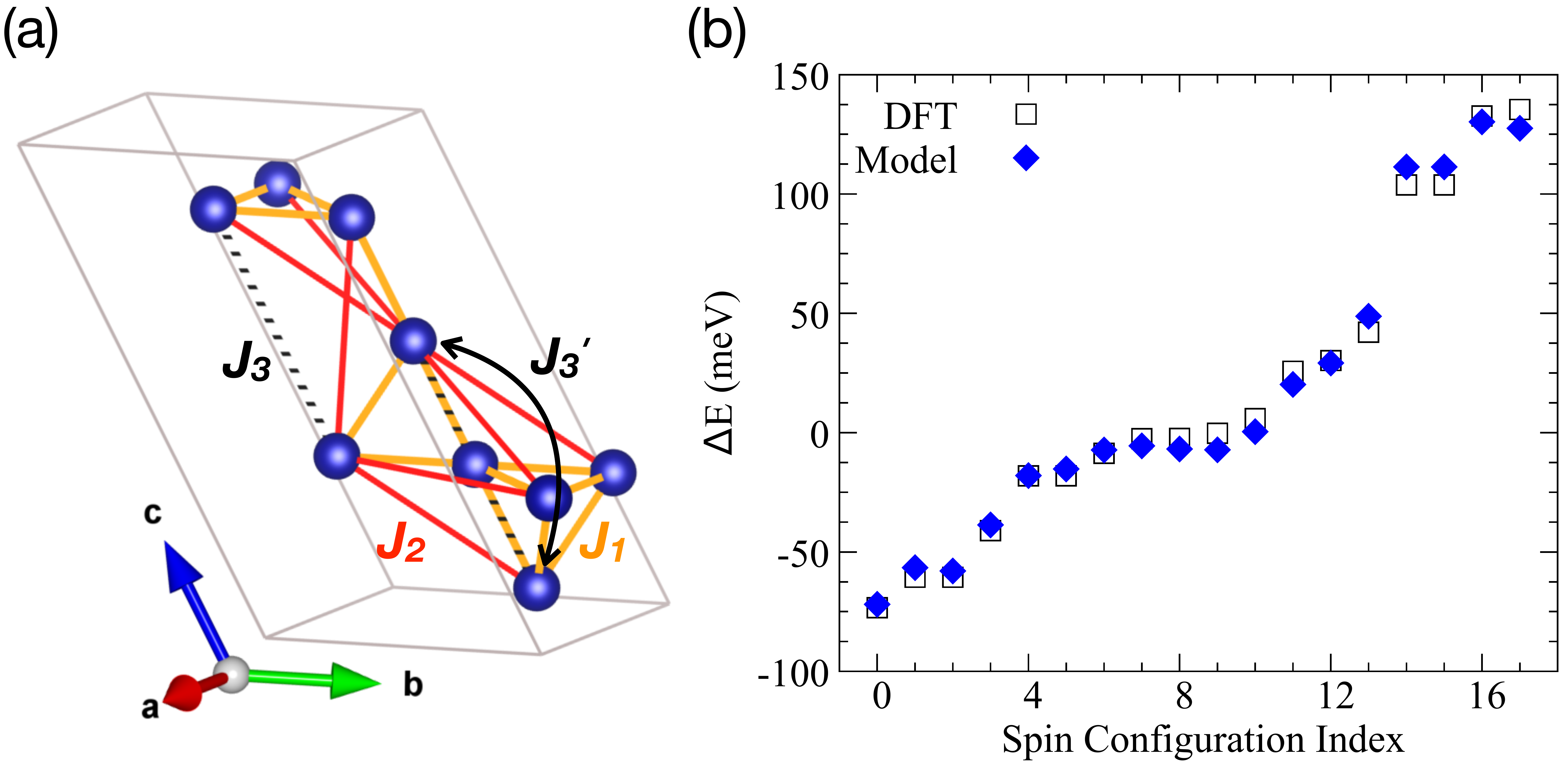}
    \caption{(a) Definition of all four magnetic-exchange interactions, first ($J_1$), second ($J_2$), and third ($J_3$ and $J_{3}^{'}$) NN, considered in this work. Co atoms are shown in blue color. Ge and O atoms are omitted for clarity. Note that $J_{3}^{'}$ passes through an intermediate Co atom (see text). (b) Fitting of the DFT (PBEsol+$U$) energy values computed for various different spin configurations in a doubled primitive cell, as shown in (a), with our model spin Hamiltonian. 
 Here, we decide to choose the PBEsol+$U$ method since it predicts better lattice parameters compared to the PBE+$U$ predictions. }
    \label{fig:jfit_fig}
\end{figure}

To get an accurate description of the low-temperature magnetic structure \ss{ experimentally reported in Ref.~\cite{pramanik2019magnetic},} we extract the values of the spin-exchange interactions ($J$'s) by mapping the DFT-computed total energies onto a Heisenberg spin Hamiltonian (Eq.~\ref{eq:hamiltonian}). 
In our spin model, we consider four exchange-interaction parameters, which correspond to the first ($J_1$), second ($J_2$), and third ($J_3$ and $J_{3}^{'}$) nearest-neighbor (NN) interactions, as shown in Fig.~\ref{fig:jfit_fig}(a). 
The first, second, and third NN interactions correspond to a Co-Co bond distance of 2.94\,\AA, 5.09\,\AA, and 5.86\,\AA, respectively. 
The third NN interaction was further divided into two categories: $J_3$ and $J_{3}^{'}$. Although both belong to the same Co-Co distance, $J_3$ connects two Co atoms located at 5.86\,\AA~distance apart without passing through any intermediate Co atom, whereas $J_{3}^{'}$ connects two Co atoms located at 5.86\,\AA~distance apart but it passes through an intermediate Co atom at the half bond distance. 
For instance, a $J_{3}^{'}$ exchange would correspond to the interaction between two Co atoms located at two adjacent TRI planes with the bond between them passing through an intermediate Co atom situated at a KGM plane [see Fig.~\ref{fig:jfit_fig}(a)].

The spin Hamiltonian reads 
\begin{equation}
\begin{aligned}
H = E_{0} + J_{1} \sum_{<ij>}^{\text{first NN}} S_{i} \cdot S_{j}  
+ J_{2} \sum_{<ij>}^{\text{second NN}} S_{i} \cdot S_{j} \\
+ J_{3} \sum_{<ij>}^{\text{third NN}} S_{i} \cdot S_{j} 
+ J_{3}^{'} \sum_{<ij>}^{\text{third NN}} S_{i} \cdot S_{j},
\end{aligned}
\label{eq:hamiltonian}
\end{equation}
where $S_{i}$ and $S_{j}$ denote the spin ordering at different Co sites, and $E_{0}$ represents a rigid shift in the total energy ($E$). 
In Fig.~\ref{fig:jfit_fig}(b), we show the fitting of the DFT+$U$ energies ($\Delta\text{E} = E-E_{0}$) computed for several distinct spin configurations in a doubled primitive cell, as shown in Fig.~\ref{fig:jfit_fig}(a), with our spin Hamiltonian described in Eq.~\ref{eq:hamiltonian}. 
The lowest-energy spin configuration corresponds to a $T_{+}\,K_{+}\,T_{-}\,K_{-}\,T_{+}\,K_{+}\,T_{-}\,K_{-}\cdots$ type AFM order, as shown in Fig.~\ref{fig:fig1struct}(b). 
This spin configuration represents a $\rm{\bf{q}}$ = ($\frac{1}{2}, \frac{1}{2}, \frac{1}{2}$) AFM order that has been experimentally observed in GCO~\cite{pramanik2019magnetic}. 
\ss{We note that all the considered spin configurations yielded gapped densities of states, shown in SM~\cite{SM}, in their converged electronic ground state. This ensured that our DFT+$U$ calculations correctly converged for all distinct spin configuration considered in this study. }

%%%%%%%%%%%%%%%%%%%%%%%%%%%%%
%%%%%%%%%%%%%%%%%%%%%%%%%%%%%
%\vspace{-0.5cm}

The best fit of data \ss{(provided in SM~\cite{SM})} yields $J_{1}S^{2}$ = $-$3.9, $J_{2}S^{2}$ = 0.7, $J_{3}S^{2}$ = 2.0, and $J_{3}^{'}S^{2}$ = 0.4 (in meV units), where positive (negative) values represent AFM (FM) magnetic interactions. We notice that the first NN exchange has a dominating FM nature, whereas all the second and third NN interactions exhibit an AFM nature, which is consistent with the recent experimental observations~\cite{pramanik2019magnetic}. 
\ss{According to the Goodenough-Anderson-Kanamory rules~\cite{morrish2001physical, Anderson1950, anderson1956ordering, Goodenough1955, Goodenough1958, Kanamori1959, Kanamori1960}, $J_{1}$ is mediated via an intermediate oxygen ion having a Co-O-Co bond angle of  $\theta=90^{0}$. Therefore, it is a superexchange interaction of FM nature. All other higher-order exchange interactions, {\it viz.,} $J_{2}$, $J_{3}$, and $J_{3}'$, are super-super AFM exchange interactions as they involve more than one ion along the exchange path.} 
These competing FM and AFM exchange interactions are primarily responsible for introducing the magnetic frustration and establishing a $\rm{\bf{q}}$ = ($\frac{1}{2}, \frac{1}{2}, \frac{1}{2}$) AFM order in GCO at low temperatures~\cite{diaz2006magnetic}.

\ss{Our theoretical findings discussed above, when combined with the experimental results reported in Ref.~\cite{pramanik2019magnetic}, provide a firm foundation for the magnetic properties of GCO. Hereafter, we focus on the lattice dynamics and vibrational properties of GCO. }

\subsection{Lattice dynamics and vibrational spectroscopy in  G\MakeLowercase{e}C\MakeLowercase{o}$_2$O$_4$}

\begin{figure}[htb!]
    \centering
    \includegraphics[width=0.84\columnwidth]{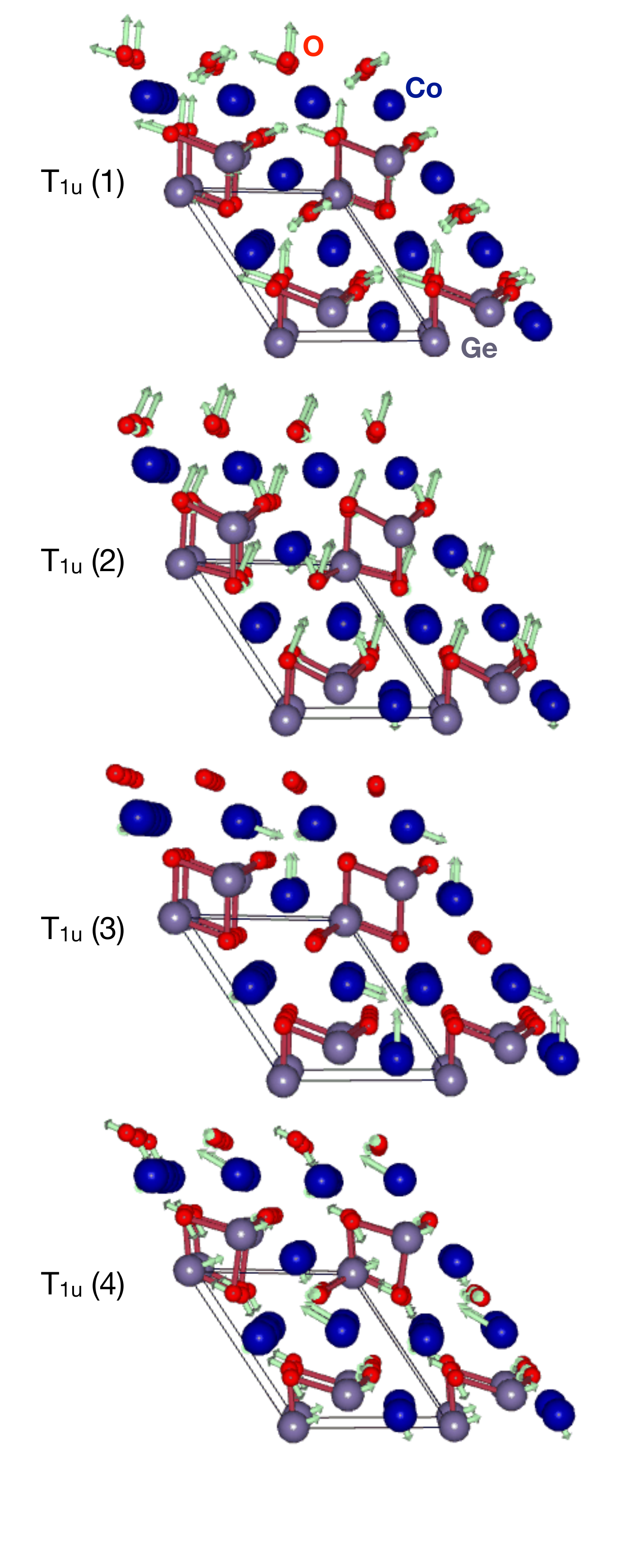}
    \caption{Atomic vibration patterns for all four IR-active phonon modes: (i) $T\textsubscript{1u}(1)$, 
    (ii) $T\textsubscript{1u}(2)$, (iii) $T\textsubscript{1u}(3)$, and (iv) $T\textsubscript{1u}(4)$. The color coding of atoms is the same as in Fig.~\ref{fig:fig1struct}(a). These modes are listed here in the order of decreasing frequency (see Table~\ref{tab:table1}).
    %Color coding of atoms is same as in Figure~\ref{fig:fig1struct}. 
    }
    \label{fig:IRmodes}
\end{figure}

\begin{figure}[htb!]
    \centering
    \includegraphics[width=0.80\columnwidth]{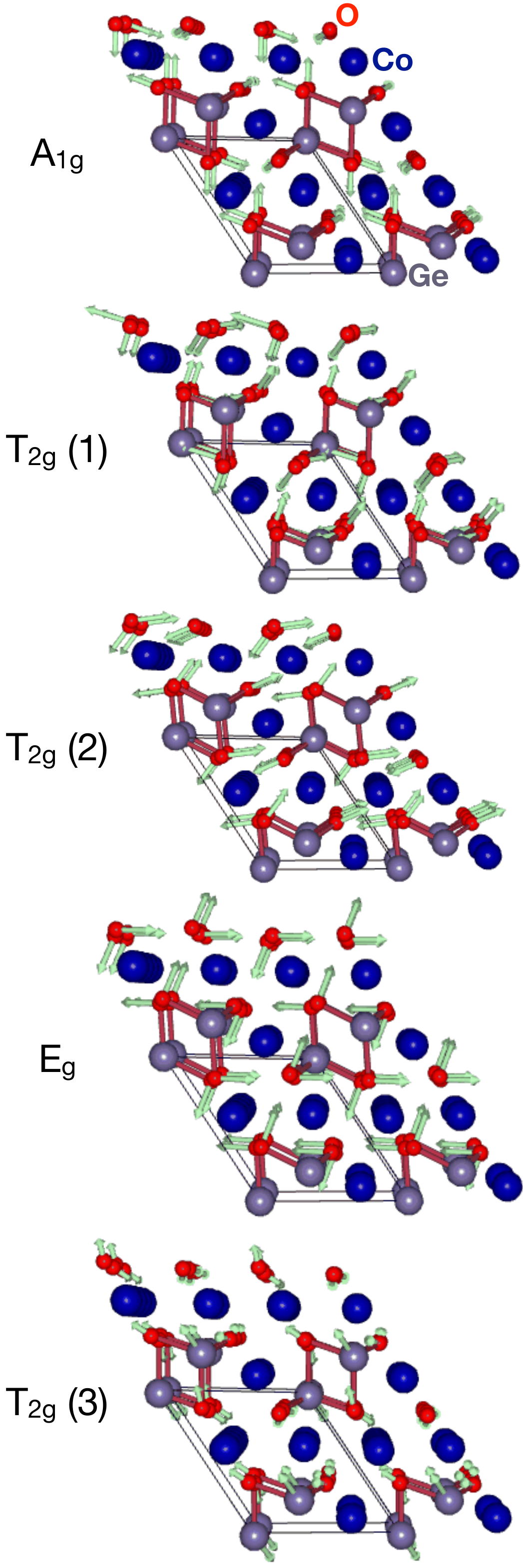}
    \caption{Atomic vibration patterns for all five Raman-active phonon modes: 
    $A$\textsubscript{1g}, $T$\textsubscript{2g}(1), $T$\textsubscript{2g}(2), 
    $E$\textsubscript{g}, and $T$\textsubscript{2g}(3). 
    %The color coding of atoms is the same as in Fig.~\ref{fig:fig1struct}(a).  
    These modes are listed here in the order of decreasing frequency (see Table~\ref{tab:table2}). 
    }
    \label{fig:RAmodes}
\end{figure}

The vibrational spectroscopy of AB\textsubscript{2}O\textsubscript{4} cubic spinels was first studied by Waldron who analyzed the phonon modes of simple ferrites (AFe\textsubscript{2}O\textsubscript{4}) using the structural primitive cell having 14 atom per cell \cite{waldron1955infrared}. 
Later, White and DeAnglis presented a group theoretical approach to analyze the Raman spectra of cubic spinels by considering the rhombohedral lattice as the smallest Bravais cell~\cite{white1967interpretation}. 
In their study, they considered the body diagonal elements consisting of two AO\textsubscript{4} and one B\textsubscript{4} tetrahedron of total 14 atoms~\cite{white1967interpretation}, as shown in Fig.~\ref{fig:fig1struct}(a). 
According to theory, the $\emph{Fd}\bar{3}\emph{m}$ space group belongs to the O\textsuperscript{7}\textsubscript{h} spectroscopic symmetry, where Ge\textsuperscript{4+}, Co\textsuperscript{2+}, and O\textsuperscript{2-} ions belong to the \emph{T}\textsubscript{d}, \emph{D}\textsubscript{3d}, and \emph{C}\textsubscript{3v}(32 \emph{e}-sites) point groups, respectively \cite{white1967interpretation}. All the
allowed optical phonon modes at the Brillouin-zone center $\Gamma$ 
(\(\overrightarrow{k} = 0\)) for each atomic displacement in the structural primitive cell can be denoted as~\cite{white1967interpretation, ChanPRB2007}:
\begin{equation}
\begin{aligned}
\Gamma\textsubscript{vib} = \emph{A}\textsubscript{1g} \oplus
2\emph{A}\textsubscript{2u} \oplus  \emph{E}\textsubscript{g} \oplus 
2\emph{E}\textsubscript{u} 
\oplus \emph{T}\textsubscript{1g} \\
\oplus\,
 4\emph{T}\textsubscript{1u} \oplus  3\emph{T}\textsubscript{2g}
\oplus 2\emph{T}\textsubscript{2u}.
\end{aligned}
\end{equation}
Out of the 39 optical phonon modes, only five modes are Raman active (\emph{A}\textsubscript{1g} $\oplus$ \emph{E}\textsubscript{g}
$\oplus$ 3\emph{T}\textsubscript{2g}),  four modes (4\emph{T}\textsubscript{1u}) are IR active, and the remaining modes are inactive in simple Raman and IR experiments. We note that the acoustic modes transform according to the $T_{1u}$ irreducible representation of the $O_{h}$ point group. The atomic vibration patterns corresponding to the IR-active modes are shown in Fig.~\ref{fig:IRmodes} and those of the Raman-active modes are shown in Fig.~\ref{fig:RAmodes}. 
These vibrational patterns, $i.e.,$ the phonon eigenvectors at $\Gamma$ depicted using green arrows, were obtained using the {\sc phonopy} package~\cite{phonopy}. 
\ss{Besides, we note that in the case of the cubic-to-tetragonal phase transition, splitting of some phonon degeneracies occurs due to the reduction in the crystal symmetry. For instance, a triply-degenerate $T_{1u}$ phonon mode splits into a doublet ($E_{u}$) and a singlet ($A_{2u}$) during the cubic to tetragonal phase transition in GCO. However, the total number of phonon modes remains the same since the cubic-to-tetragonal phase transition is primarily driven by a zone-center $\Gamma_{3}^{+}$ mode.   
}

%%%%%%%%%%%%%%%%%%%%%%%%%%%%%
%%%%%%%%%%%%%%%%%%%%%%%%%%%%%

%%%%%%%%%%%%%%%%%%%%%%%%%%%%%
%%%%%%%%%%%%%%%%%%%%%%%%%%%%%

As mentioned earlier, the magnetic structure of GCO is quite complex due to the $\rm{\bf{q}}$ = ($\frac{1}{2}, \frac{1}{2}, \frac{1}{2}$) AFM ordering,  and a first-principles DFT+$U$ calculation of the full phonon dispersion for the actual magnetic cell would be computationally very demanding. However, DFT+$U$ calculation for the structural primitive cell (14 atoms/cell) considering various different spin configurations can provide useful insights about the Raman/IR-active phonon modes at the zone-center $\Gamma$ (which is required for this study), and the strength of the spin-phonon coupling in GCO. 

To simulate the high-temperature paramagnetic phonon frequencies (at the infinite temperature limit of spin fluctuations), we follow the method proposed by Kumar-Fennie-Rabe for magnetic spinels~\cite{KFR2012}. 
In this method, we take the statistical average of the interatomic force constants calculated for all the possible spin configurations such that each Co-Co bond has an equal fraction of parallel and antiparallel spins. 
This method assumes that the time scale of phonons is much longer compared to the spin fluctuations, and spins in the paramagnetic phase are not correlated, 
which are reasonable approximations at the high-temperature limit. 
In the case of GCO, we have 4 magnetic Co atoms yielding a total of $2^{4}$ (=16) collinear spin configurations, which can be reduced to 8 spin configurations using the time-reversal symmetry. 
A further consideration of the cubic crystal symmetry reduces the total number of non-equivalent spin configurations to three, which are: ${++++}$, ${++--}$, and ${+---}$, each with a statistical weight of $\frac{1}{8}$, $\frac{3}{8}$, and $\frac{1}{2}$, respectively. Here $+/-$ denotes the up/down spin moment at each Co site. 
Thus computed phonon frequencies for the IR-active and Raman-active 
modes are given in Table~\ref{tab:table1} and Table~\ref{tab:table2}. 

Owing to the fact that the PBEsol functional  describes the lattice parameters and bond lengths in GCO better compared to the PBE functional, we find that the PBEsol predicted phonon frequencies are in better agreement with the experimental data as compared to the PBE predictions.

%\vspace{-0.2cm} 
\subsubsection{{\bf{IR-active modes}}}
\vspace{-0.2cm}

\begin{figure}[htb]
    \centering
    \includegraphics[trim=1.7cm 0.2cm 1cm 2cm, clip=true,scale=0.37]{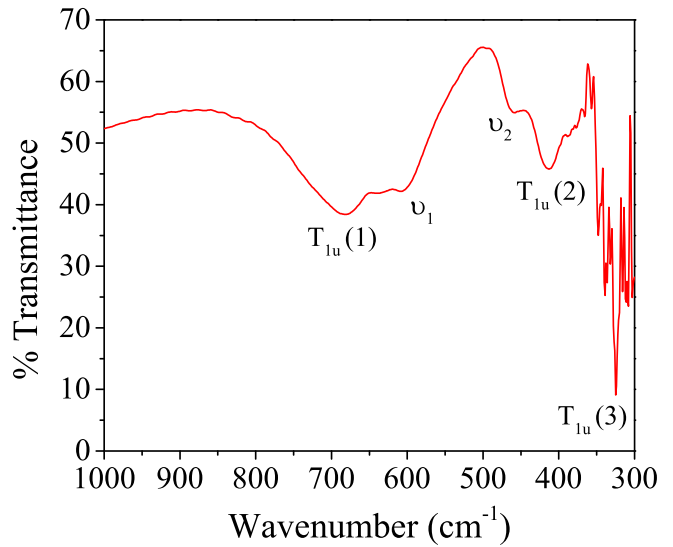}
    \caption{Fourier-transform infrared spectrum of GCO polycrystalline sample recorded at room temperature. \ss{The DFT+$U$ simulated IR spectrum is given in the SM~\cite{SM}. }}
    \label{fig:irspectra}
\end{figure}

The frequency of the four allowed IR-active modes in GCO along with \ss{those} for some other normal spinels \ss{are} listed in Table~\ref{tab:table1}. 
The Fourier-transform infrared (FTIR) spectrum of GCO recorded at 300\,K in the transmission mode, shown in Fig.~\ref{fig:irspectra}, 
displays the observation of $T$\textsubscript{1u}(1), $T$\textsubscript{1u}(2), and
$T$\textsubscript{1u}(3) modes at frequencies 680, 413, and 325 \(\text{cm}^{-1}\), respectively, which are in decent agreement with our DFT+$U$ calculated frequencies. 
Since the experimental limitations did not allow us to measure modes below
300 \(\text{cm}^{- 1}\), the $T$\textsubscript{1u}(4) mode predicted to occur at
189 \(\text{cm}^{- 1}\ \) (see Table~\ref{tab:table1}) could not be observed. 
However, the predicted frequency of the $T$\textsubscript{1u}(4) mode is in good agreement with the experimental data (186 \(\text{cm}^{- 1}\ \)) reported by Preudhomme and Tarte~\cite{preudhomme1972infrared}. 
Overall there is a good agreement between the observed and predicted values for the IR-active modes at room temperature, \ss{as shown in Figure~\ref{fig:compare_ir_raman}. }

In addition to the above listed IR-active modes, Fig.~\ref{fig:irspectra}
shows the observation of two satellite modes at 608 and
459 \(\text{cm}^{- 1}\), marked as \(v_{1}\ \)and
\(v_{2}\), respectively.  
Although crystal symmetry allows the observation of only four $T$\textsubscript{1u} modes, these additional satellite modes are likely occurring from the  splitting of the $T$\textsubscript{1u} modes due to the induced local electric fields~\cite{ChanPRB2007}. 
The presence of any impurity or crystallite domains \ss{in a powder sample} breaks the local crystal symmetry distorting the local potential, which in turn relaxes the selection rules governing the observation of the allowed IR-active modes, and it may lead to the appearance of the satellite modes in the IR spectrum. Such satellite modes have been previously observed in lithium-cobalt oxides~\cite{BURBA2009248, Ahamed_2020}. 

Our DFT+$U$ calculations predict moderate spin-phonon coupling in the IR-active T$_{1u}$ modes of GCO. We notice that each triply-degenerate T$_{1u}$ mode of the $O_h$ point group splits into two modes, one doublet and one singlet, when the magnetic symmetry is changed from FM to AFM, which is consistent with the work of Wysocki and Birol~\cite{WysockiBirol2016}. The magnitude of the frequency splitting between the doublet and singlet modes ($\Delta\omega_{ds}$) provides a good qualitative estimate of the strength of the spin-phonon coupling in magnetic spinels~\cite{FenniePRL2006, KFR2012, ChanPRB2007, WysockiBirol2016}. In case of GCO, the PBEsol+$U$ (PBE+$U$) calculated $\Delta\omega_{ds}$ is 1 (1), 4 (2), 6 (10), 2 (2) \(\text{cm}^{- 1}\) for the 
\emph{T}\textsubscript{1u}(1), \emph{T}\textsubscript{1u}(2), \emph{T}\textsubscript{1u}(3), and \emph{T}\textsubscript{1u}(4) modes, respectively. 
These values are consistent with the previously reported data on other magnetic spinels~\cite{FenniePRL2006, KFR2012, ChanPRB2007, WysockiBirol2016}. The maximum frequency splitting is predicted for the \emph{T}\textsubscript{1u}(3) mode, which is evident since the \emph{T}\textsubscript{1u}(3) mode involves the vibration of the magnetic Co sites, as shown in Fig.~\ref{fig:IRmodes}. 
An experimental validation of the aforementioned frequency-splitting values requires low temperature IR measurements, which, unfortunately, could not be carried out because of the limitations of our experimental facilities.

The high frequency IR-active modes \emph{T}\textsubscript{1u}(1) and \emph{T}\textsubscript{1u}(2), as shown in Fig.~\ref{fig:IRmodes}, involve the symmetric and asymmetric bending of oxygen ions present at the tetrahedral and octahedral sites, whereas the low frequency IR-active modes, \emph{T}\textsubscript{1u}(3) and \emph{T}\textsubscript{1u}(4), are associated with the vibrations of the relatively heavier Ge and Co ions situated at the tetrahedral and octahedral sites, respectively. 
Generally, the frequency of a mode varies as \(\sqrt{k/m},\) where $k$ is the stiffness constant of the bond and $m$ is the effective mass of the associated ions. 
From the magnitudes of the four IR-active modes for various spinels listed in Table~\ref{tab:table1},
one can argue that $T$\textsubscript{1u}(1) and \emph{T}\textsubscript{1u}(2) modes are due to the vibrations of the tetrahedral group (\(\text{GeO}_{4}\) or \(\text{SiO}_{4}\)) whereas $T$\textsubscript{1u}(3) and 
$T$\textsubscript{1u}(4) also involve the vibrations of the octahedral group (\(\text{MgO}_{6}\) and \(\text{CoO}_{6}\)). 
Our reasoning is as follows: 
When Co in GeCo\textsubscript{2}O\textsubscript{4} is replaced by
lighter Mg in GeMg\textsubscript{2}O\textsubscript{4}, there is
about 50\% increase in the frequencies of $T$\textsubscript{1u}(3) and $T$\textsubscript{1u}(4) modes, whereas the increase in the frequencies of the $T$\textsubscript{1u}(1) and $T$\textsubscript{1u}(2) modes is only a few percent. When lighter Si in SiCo\textsubscript{2}O\textsubscript{4} replaces heavier Ge in GeCo\textsubscript{2}O\textsubscript{4}, then the frequencies
of the $T$\textsubscript{1u}(1)and $T$\textsubscript{1u}(2) modes in SiCo\textsubscript{2}O\textsubscript{4} goes up by about 25\%, whereas changes 
in the $T$\textsubscript{1u}(3) and $T$\textsubscript{1u}(4) mode frequencies is only about 5\%. Therefore, $T$\textsubscript{1u}(1) and $T$\textsubscript{1u}(2) modes primarily represent the
vibrations of the tetrahedral group, while $T$\textsubscript{1u}(3) and $T$\textsubscript{1u}(4) modes represent the
vibrations of the octahedral group. 
This qualitative description is consistent with the schematic phonon eigenvectors plot shown in Fig.~\ref{fig:IRmodes}.

\subsubsection{{\bf{Raman-active modes }}}
\vspace{-0.2cm}
The frequencies of the Raman-active modes 
in GCO (at 300\,K) are listed in Table~\ref{tab:table2} along with their 
calculated values. As done in Table~\ref{tab:table1} for the IR modes, we have also 
listed the frequencies of the Raman-active modes in Table~\ref{tab:table2} reported for several other isostructural spinels \emph{e.g.,}  SiCo\textsubscript{2}O\textsubscript{4},
GeMg\textsubscript{2}O\textsubscript{4}, 
MgTi\textsubscript{2}O\textsubscript{4}, and 
SiMg\textsubscript{2}O\textsubscript{4}. 
Our observed values of the frequencies of $A$\textsubscript{1g}, $T$\textsubscript{2g}(1), and  $E$\textsubscript{g} modes in GCO are nearly identical to those
reported by Koningstein \emph{et al.}~\cite{koningstein1972light}.
The frequency of the $T$\textsubscript{2g}(2) mode in GCO is reported for the first time in this work. 
Our DFT+$U$ calculated phonon frequencies of Raman-active modes are in good agreement with the experimental observations. The  $T$\textsubscript{2g}(3) mode could not be detected in our experiments since this mode is predicted to occur below the lowest frequency of our Raman measurements. However, the predicted frequency of the  $T$\textsubscript{2g}(3) mode is consistent with that of reported values for other isostructural spinel oxides (see Table~\ref{tab:table2}).

\begin{figure}[htb!]{
    \centering
    \includegraphics[trim=2cm 30cm 1cm 9.5cm,clip=true,scale=0.115]{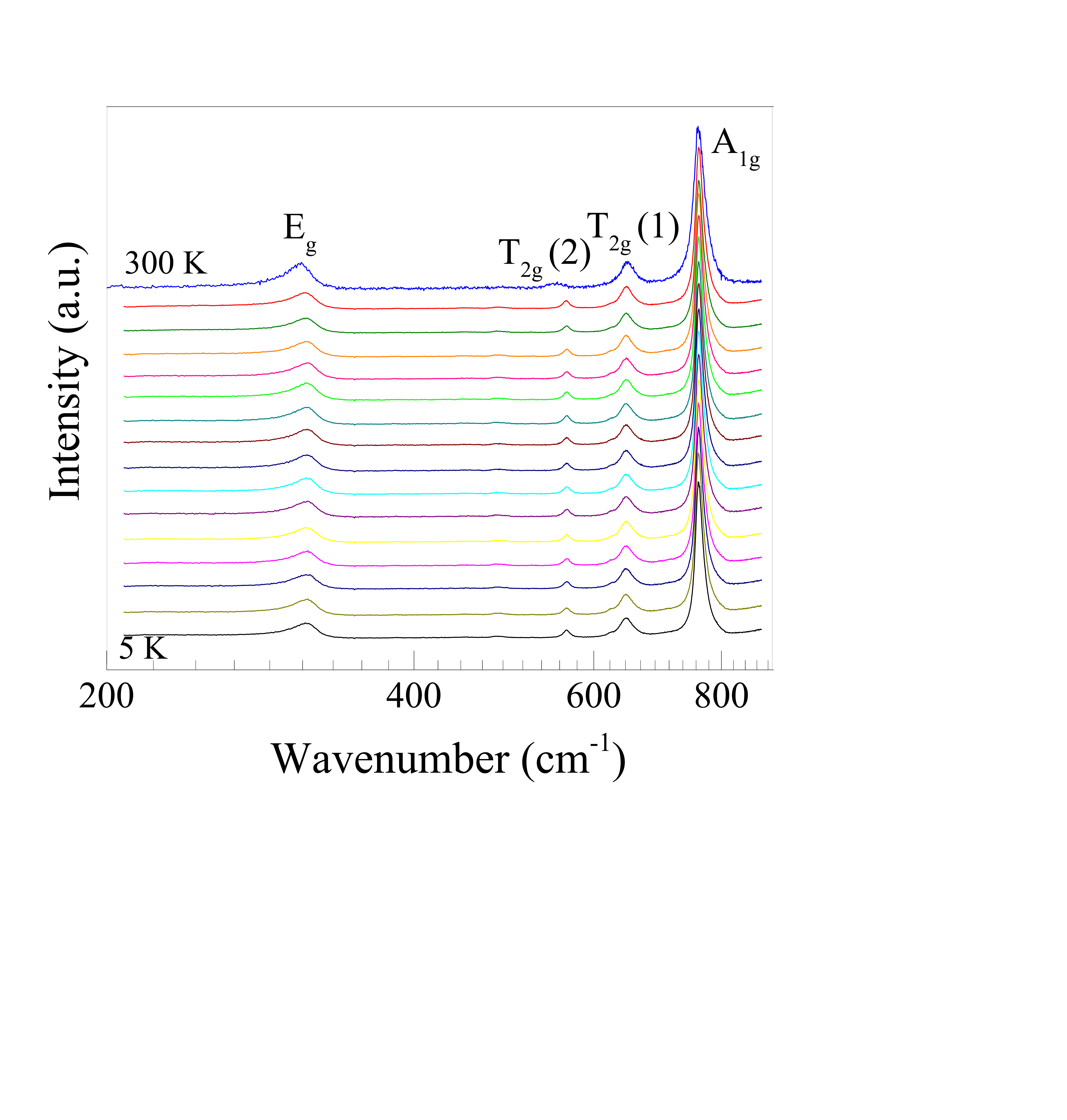}
    \caption{Raman spectra of GCO recorded at temperatures T = 5, 10, 12, 14,  18, 20, 21, 22, 24, 26, 30, 40, 60, 80, 100, and 300\,K. 
    }  
     \label{fig:raman_temp5_300} }
\end{figure}

Our calculations reveal that the strength of the spin-phonon coupling is the largest for the $T$\textsubscript{2g}(3) mode since this mode is associated with the vibration of the heavy cations. The values of the frequency splitting $\Delta\omega_{ds}$ for the triply-degenerate $T$\textsubscript{2g}(1), $T$\textsubscript{2g}(2), and $T$\textsubscript{2g}(3) modes are 3 (1), 2 (2), 5 (3) \(\text{cm}^{- 1}\), respectively, as obtained using the PBEsol+$U$ (PBE+$U$) method. 
We note that these values are in the same range of the observed frequency shifts of the associated Raman peaks at \emph{T\textsubscript{N}}, as discussed below. 

To better understand the Raman modes in GCO, a systematic comparison of their frequencies with those reported in SiCo\textsubscript{2}O\textsubscript{4},
GeMg\textsubscript{2}O\textsubscript{4}, 
SiMg\textsubscript{2}O\textsubscript{4}, and MgTi\textsubscript{2}O\textsubscript{4} are listed in Table~\ref{tab:table2}. Comparing
 SiCo\textsubscript{2}O\textsubscript{4} with
GeCo\textsubscript{2}O\textsubscript{4} for which lighter Si atom
replaces heavier Ge atom at the tetrahedral site, the frequencies of the $A$\textsubscript{1g}, $E$\textsubscript{g}, and $T$\textsubscript{2g}(1) modes in
SiCo\textsubscript{2}O\textsubscript{4} are increased by about 10--20\%. 
This suggests that these modes likely involve some motion of the
tetrahedral cation in addition to the O atoms.
This is further confirmed by comparing the mode frequencies of
GeMg\textsubscript{2}O\textsubscript{4} with those in
SiMg\textsubscript{2}O\textsubscript{4} where the frequencies of
$A$\textsubscript{1g},$E$\textsubscript{g}, and $T$\textsubscript{2g}(1) modes in
SiMg\textsubscript{2}O\textsubscript{4} are higher by about 10--20\%.
For the  $T$\textsubscript{2g}(2)  mode, the observed differences in the
frequencies for GCO {\it vis-a-vis} SiCo\textsubscript{2}O\textsubscript{4},
GeMg\textsubscript{2}O\textsubscript{4}, and
SiMg\textsubscript{2}O\textsubscript{4} do not show a systematic pattern. 
%To understand the role of cations at the octahedral site on the Raman modes, mode frequencies in 
To further understand the role of the Co-O octahedra on the Raman modes, mode frequencies in GeMg\textsubscript{2}O\textsubscript{4} and
GeCo\textsubscript{2}O\textsubscript{4} are compared for which the lighter Mg replaces the heavier Co. 
In this case, the frequencies of the $A$\textsubscript{1g} and $T$\textsubscript{2g}(1) modes are increased by about 2\% only. 
However, the frequency of the $E$\textsubscript{g} mode in GeMg\textsubscript{2}O\textsubscript{4} is enhanced by about 11\%. 
This suggests that the $E$\textsubscript{g} mode also involves some vibrations of the cations on the octahedral site. 
In summary, for GCO, the $A$\textsubscript{1g} and $T$\textsubscript{2g}(1) modes involve some vibrations of Ge at the tetrahedral site in addition to the vibrations of the O atoms, whereas for the $E$\textsubscript{g} modes, the vibrations of GeO\textsubscript{4} and CoO\textsubscript{6} are also involved.

%%%%%%%%%%%%%%Table1%%%%%%%%%%%%%
\begin{table*}[htb]
%\arraystretch{1.5}
\begin{center}
\caption{List of IR-active modes and 
their frequencies (in cm$^{- 1}$) at room temperature for several cubic spinels}
\begin{tabular}{|c|c|c|c|c|c|}
\hline
 & $T$\textsubscript{1u}(1)  &  $T$\textsubscript{1u}(2) & $T$\textsubscript{1u}(3) & $T$\textsubscript{1u}(4) & Reference \\
\hline
%GeCo\textsubscript{2}O\textsubscript{4} & 680.2 & 413.4 & 325 &   & This work (Experiment) \\
GeCo\textsubscript{2}O\textsubscript{4} & 680 & 413 & 325 &    & This work (Experiment) \\
\hline
GeCo\textsubscript{2}O\textsubscript{4} & $^{\#}$\,640 & 407 & 312 & 189  & This work (Calculation) \\
 & *\,(615) & (379) & (294) & (168) &  \footnotesize{$^{\#}$\,PBEsol+$U$; *\,(PBE+$U$)} \\
\hline
GeCo\textsubscript{2}O\textsubscript{4} & 679 & 427 & 321 & 186  & \cite{preudhomme1972infrared} \\
\hline
GeNi\textsubscript{2}O\textsubscript{4} & 690 & 453 & 335 & 199  & \cite{preudhomme1972infrared} \\
\hline
GeMg\textsubscript{2}O\textsubscript{4} & 694 & 450 & 485 & 274  & \cite{preudhomme1972infrared} \\
\hline
SiCo\textsubscript{2}O\textsubscript{4} & 815 & 504 & 354 & 161  & \cite{kushwaha2018vibrational} \\
\hline
SiMg\textsubscript{2}O\textsubscript{4} & 834 & 547 & 444 & 348  & \cite{kushwaha2018vibrational} \\
\hline
\end{tabular}
\label{tab:table1}
\end{center}
%\footnotesize{*PBE+$U$; \# PBEsol+$U$}
\end{table*}

\begin{table*}[htb]
\begin{center}
\caption{List of Raman-active phonon modes and their frequencies (in cm$^{- 1}$) at room temperature for several cubic spinels}
\begin{tabular}{|c|c|c|c|c|c|c|}
\hline
 & $A$\textsubscript{1g}  & $T$\textsubscript{2g}(1) &   $T$\textsubscript{2g}(2) & $E$\textsubscript{g} & $T$\textsubscript{2g}(3) &  Reference \\
\hline
GeCo\textsubscript{2}O\textsubscript{4} & 760 &  647 & 550 & 308   &  & This work (Experiment) \\
\hline
GeCo\textsubscript{2}O\textsubscript{4} & $^{\#}$\,720 & 649  & 475 & 323 & 204 & This work (Calculation)  \\
 & *\,(695) & (610)  & (461) & (311) & (203) & \footnotesize{$^{\#}$\,PBEsol+$U$; *\,(PBE+$U$)} \\
\hline
GeCo\textsubscript{2}O\textsubscript{4} &757 & 643 &  & 302   &  & \cite{koningstein1972light} \\
\hline
SiCo\textsubscript{2}O\textsubscript{4} & 833  & 788 & 521 & 373 & 270 & \cite{kushwaha2018vibrational} \\
\hline
GeMg\textsubscript{2}O\textsubscript{4} & 777 & 669  & 520 & 341 & 213 & \cite{ross1987mg} \\
\hline
SiMg\textsubscript{2}O\textsubscript{4} & 834 & 798  & 599 & 373 & 300 & \cite{kushwaha2018vibrational} \\
\hline
MgTi\textsubscript{2}O\textsubscript{4} &628 & 493  & 335 & 448 &  & \cite{popovic2003phonon} \\
\hline
\end{tabular}
\label{tab:table2}
\end{center}
\end{table*}

\begin{figure}[h!]
    \centering
    \includegraphics[width=0.96\columnwidth]{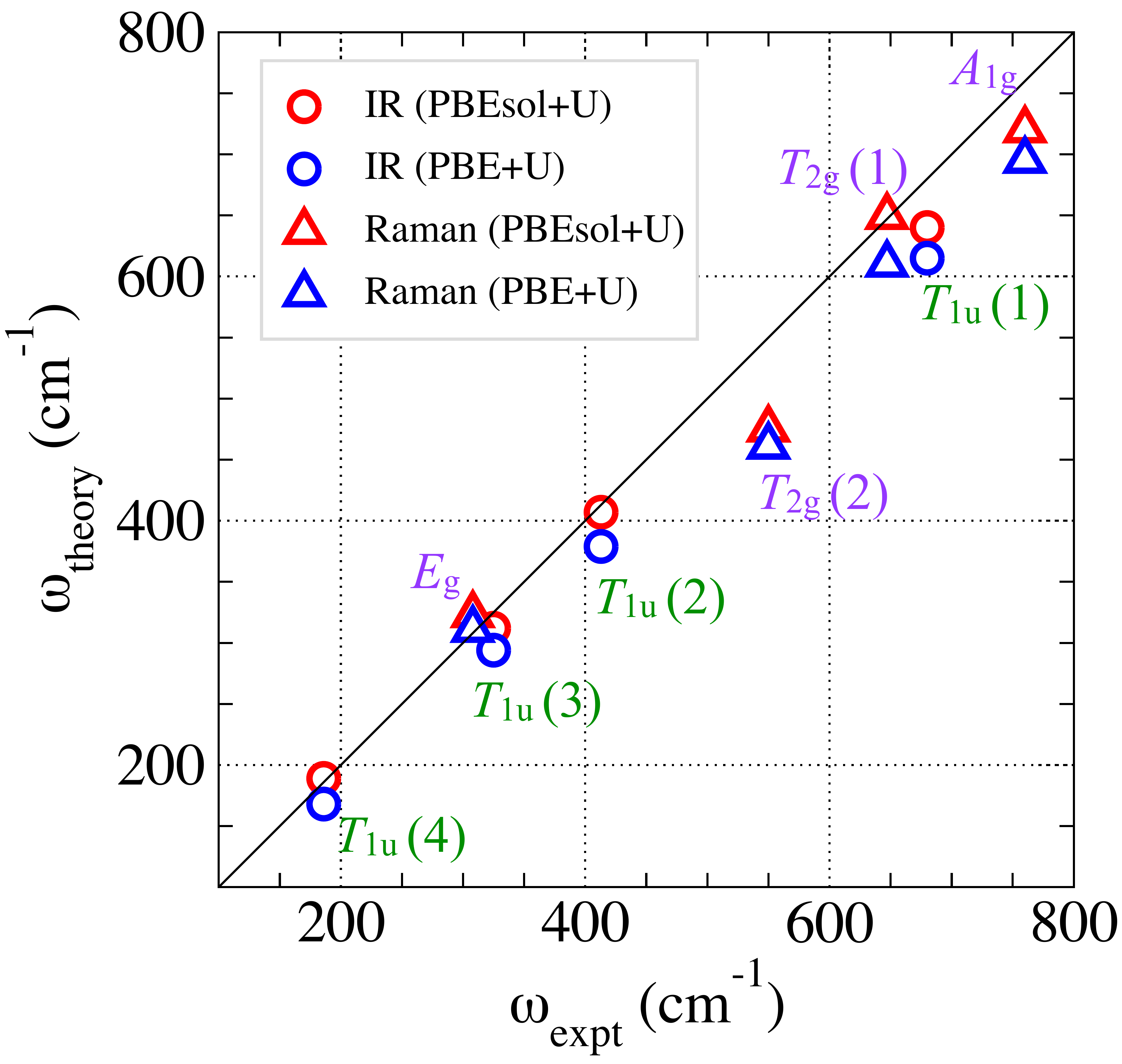}
    \caption{ \ss{Comparison of the DFT+$U$ predicted phonon frequencies ($\omega$\textsubscript{theory}) for the simulated paramagnetic phase with the experimentally measured frequencies at 300\,K ($\omega$\textsubscript{expt}) for the IR and Raman-active modes. The data plotted in this figure were obtained from Table~\ref{tab:table1} and~\ref{tab:table2}. }
    }
    \label{fig:compare_ir_raman}
\end{figure}

\ss{In Figure~\ref{fig:compare_ir_raman} we compare the DFT+$U$ predicted phonon frequencies calculated for the paramagnetic phase using the statistical averaging method, as mentioned above, with the experimental data recorded at 300\,K from the IR and Raman measurements. The experimental frequency of the $T$\textsubscript{1u}(4) mode was obtained from Ref.~\cite{preudhomme1972infrared}. We observe a good agreement between theory and experiment. In particular, the PBEsol+$U$ predicted frequencies are in better agreement with the experimental data compared to the PBE+$U$ predictions.  }

%%%%%%%%%%%%%%%%%%%%%%%%%%%%%
%%%%%%%%%%%%%%%%%%%%%%%%%%%%%

\subsubsection{{\bf{Temperature dependence of the Raman-active modes}}}
\vspace{-0.2cm}

\begin{figure}[htb!]
    \centering
   \includegraphics[trim=3cm 0.7cm 2cm 1cm, clip=true,scale=0.6]{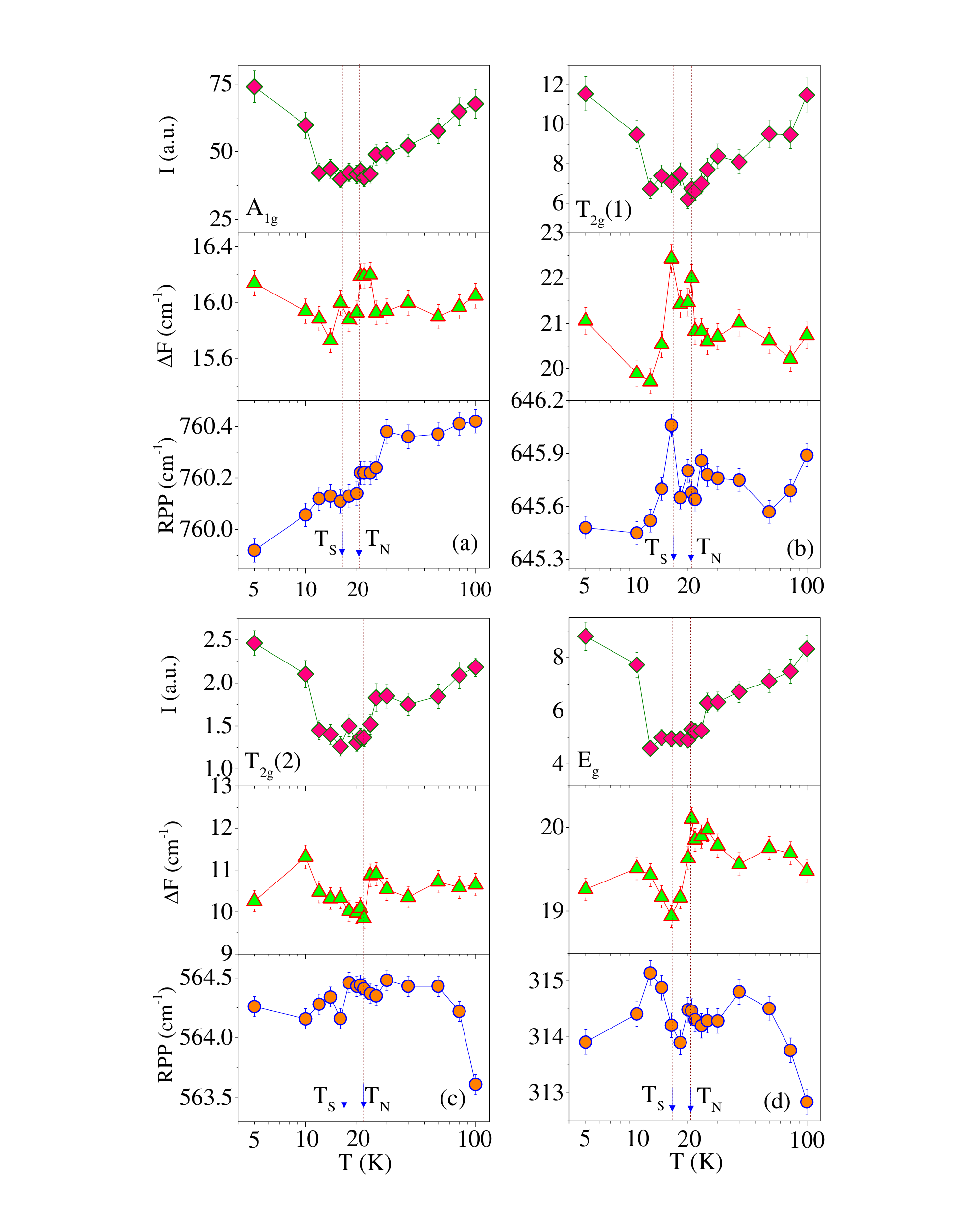}
      \caption{The temperature dependence of Raman intensity (I$\times$10$^{4}$), 
      fullwidth at half maximum ($\Delta F$), and Raman-peak position (RPP) for the $A$\textsubscript{1g}, $T$\textsubscript{2g}(1), $T$\textsubscript{2g}(2), and $E$\textsubscript{g} Raman active modes. The lines connecting the data points are visual guides. The \emph{T}\textsubscript{N} and \emph{T}\textsubscript{S} mark the transition temperatures corresponding to the antiferromagnetic and the cubic-to-tetragonal structural phase transitions, respectively.
            } 
    \label{fig:raman_peaktemp}
\end{figure}

A brief summary of the temperature dependence of the structural
properties of GCO is first presented in order to place the data on the
Raman-active modes in proper context. Using x-ray synchrotron data on a
polycrystal GCO, Barton \emph{et al}.~\cite{barton2014structural} determined changes in the lattice parameters and Co-O and Ge-O bond lengths as a function of
temperature including the regions around \emph{\emph{T}\textsubscript{N}} = 21
K and \emph{T\textsubscript{S}} = 16 K. For \emph{T} \textless{}
\emph{T\textsubscript{S}}, the crystal symmetry changes from cubic to
tetragonal with c/a \textgreater{} 1 with degree of tetragonality
increasing with decreasing \emph{T}. An elongation of the
CoO\textsubscript{6} octahedron is observed below the 
\emph{T\textsubscript{S}} as the Co-O bond length = 2.09\,\AA~above
\emph{\emph{T}\textsubscript{N}} increases to 2.13\,\AA~along the c-axis but
decreases to 2.07\,\AA~normal to c-axis for \emph{T \textless{}
T\textsubscript{S}}. However, there is no change in the Ge-O bond length
in the GeO\textsubscript{4} tetrahedron as the symmetry changes from the
cubic to tetragonal phase below the \emph{T\textsubscript{S}}. Considering
these results, changes around \emph{T\textsubscript{S}} should be
expected in the Raman and the IR-active modes which involve vibrations of the
atoms in the CoO\textsubscript{6} octahedron. 

The structural transition at low temperature is expected due to the possible Jahn-Teller distortions and spin-orbit coupling effects in the 3$d^7$ state of Co$^{2+}$, in which the spin degeneracy is lifted due to the stabilization of the $t_{2g}$ orbitals. 
Following our earlier discussion on the comparison of the Raman-active modes for different spinels listed in Table~\ref{tab:table2}, significant changes around
\emph{T\textsubscript{S}} should be expected for the $E$\textsubscript{g} mode.
Another relevant and important results from the paper by Barton {\it et al.}~\cite{barton2014structural} is the presence of the magneto-dielectric coupling, 
which is  evident from the fitting of the temperature-dependent dielectric constant data of GCO with the Barrett equation for \emph{T \textgreater{} \emph{T}\textsubscript{N}} (similar to previous reports on MnO and MnF\textsubscript{2} \cite{seehra1981dielectric,seehra1984anomalous,seehra1986effect})
yielding 339\,cm$^{-1}$ as the frequency of the coupling
mode. This frequency is close to that of the \emph{E\textsubscript{g}}
mode determined in this work.

Keeping the above comments in mind, the Raman spectra of GCO recorded at
various temperatures between 5 and 300\,K are shown in Fig.~\ref{fig:raman_temp5_300} with each line identified with one of the five Raman-active modes. For each line, except
the $T$\textsubscript{2g}(3) mode whose intensity is too weak for accurate measurements, we measured its position, full width at half maximum
(FWHM) and line intensity (area under the peak), 
and plotted these quantities as a function of temperature in Fig.~\ref{fig:raman_peaktemp}. The
positions of \emph{T}\textsubscript{N} = 21\,K and
\emph{T}\textsubscript{S} = 16\,K are also marked by vertical dashed 
lines in these plots. 
Qualitative interpretations of these results are presented below.

\begin{figure}[htb!]
    \centering
     \includegraphics[trim=0.2cm 0cm 0cm 1.8cm, clip=true,scale=0.36]{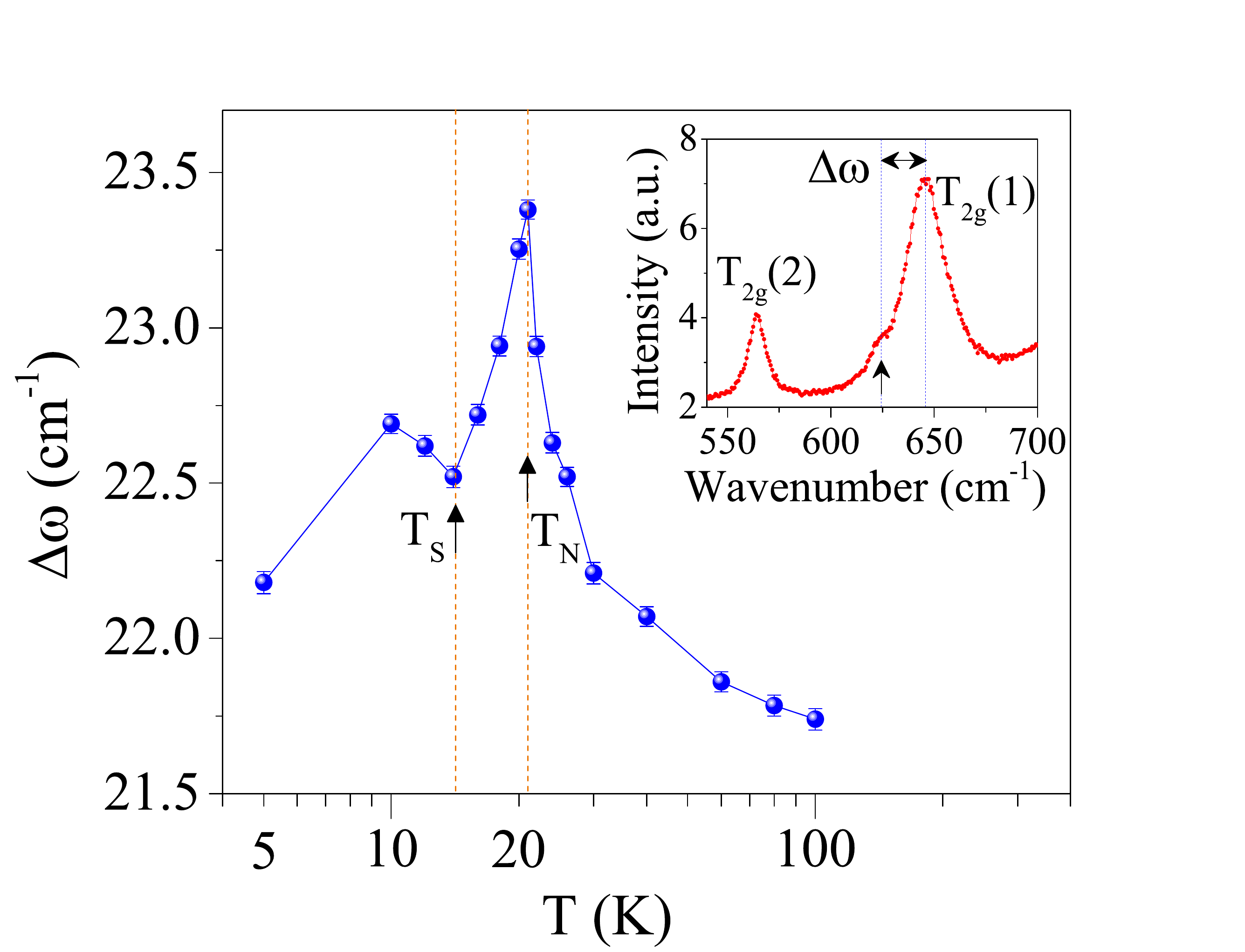}
    \caption{The temperature dependence of the line separation $\Delta \omega$ of the low-frequency shoulder from the position of the $T$\textsubscript{2g}(1) peak. Inset marks the shoulder appearing on the low-frequency side of the $T$\textsubscript{2g}(1) peak. }
    \label{fig:temp_t2g1}
\end{figure}

A detailed examination of the plots shown in Fig.~\ref{fig:raman_peaktemp} reveals some interesting features. First, for all the four observed Raman modes {\it viz.,}  $A$\textsubscript{1g}, $E$\textsubscript{g}, $T$\textsubscript{2g}(1), and
$T$\textsubscript{2g}(2), the intensity of the Raman lines increases with 
decreasing temperature below \emph{T}\textsubscript{S}, which is somewhat similar to the variation of the order parameter. 
According to the Suzuki and Kamimura theory~\cite{suzuki1973theory} for the spin-dependent Raman scattering, the magnetic order significantly influences the phonon Raman efficiency through the dependence of the optical dipole transitions on the relative orientation of the adjacent spins. 
Generally, the temperature dependence of the integrated Raman intensity is proportional to the nearest neighbor spin correlation function~\cite{balkanski1987magnetic}. 
Also, the emergence of an AFM order below the $T_N$ enhances the Raman intensity due to the Brillouin-zone folding since the magnetic unit cell would be doubled in size compared to the structural unit cell~\cite{suzuki1973theory, balkanski1987magnetic}. 
As a result, the Raman intensity always enhances below the magnetic transition in both FM and AFM systems. 

The second noticeable effect is the dramatic changes observed in the FWHM for the
$T$\textsubscript{2g}(1), $T$\textsubscript{2g}(2) and \emph{E}\textsubscript{g}
modes between \emph{T}\textsubscript{N} and \emph{T}\textsubscript{S} along with weaker anomalies in the line positions of these modes. 
As argued earlier based on the comparison 
with data on other spinels, significant changes due to the structural
transition at \emph{T}\textsubscript{S} were expected in the line
parameters of the \emph{E}\textsubscript{g} mode.
Results presented in Fig.~\ref{fig:raman_peaktemp} show that the 
effects of magnetic ordering at \emph{T}\textsubscript{N} and structural transition at
\emph{T}\textsubscript{S} for the $T$\textsubscript{2g}(1), $T$\textsubscript{2g}(2) 
and \emph{E}\textsubscript{g} modes are significant. 

The Raman linewidth is supposed to decrease with decreasing temperature since the phonon scattering usually gets suppressed at low temperatures. 
As it can be noticed from Fig.~\ref{fig:raman_peaktemp}, the FWHM ($\Delta F$) is indeed decreasing below \emph{T}\textsubscript{N} until \emph{T}\textsubscript{S}, which clearly indicates that the structural transition is independent of the magnetic transition.  
Also, for the only case of T \textless \emph{T}\textsubscript{S}, there is (roughly) an overall increase in the FWHM of all four Raman-active modes. 
This could be associated to the cubic-to-tetragonal structural distortion occurring at \emph{T}\textsubscript{S} since this distortion could lift the degeneracy of the degenerate Raman-active modes, with the exception of the  nondegenerate \emph{A}\textsubscript{g} mode. It is possible that the distortion-split modes are not showing up distinctly in our Raman measurements due to their smaller magnitude of the frequency shift, however, they may form a convoluted peak with a larger FWHM. 
Another possible explanation could be related to the local structural disorder driven by the randomly distributed Ge atoms, which may cause an increase in the linewidth at \emph{T} $<$  \emph{T}\textsubscript{S}.

Another noteworthy feature evident from the Raman spectra at low temperatures is the separation of a shoulder, marked by an arrow in the inset of Fig.~\ref{fig:temp_t2g1}, on the low-frequency side of the $T$\textsubscript{2g}(1) line. 
The origin of this shoulder is not yet well understood. However, we think it could be attributed to the magnon-induced excitations~\cite{Zhang_NatComm2016}.  
In Fig.~\ref{fig:temp_t2g1}, we plot the temperature dependence of the frequency shift of this shoulder $\Delta \omega$ from the  $T$\textsubscript{2g}(1) line. 
We note that $\Delta \omega$ increases with lowering temperature and attains a maximum value at \emph{T}\textsubscript{N}. With a further decrease in temperature (\emph{T}\textsubscript{S} $<$ \emph{T} $<$ \emph{T}\textsubscript{N}), $\Delta \omega$ starts decreasing, and it shows an upturn at \emph{T}\textsubscript{S}. 

Such a temperature dependence of $\Delta \omega$ implies the  presence of two distinct phase transitions, one magnetic and another structural, in GCO, thus, validating the claim of Barton \emph{et al.}~\cite{barton2014structural} that the structural phase transition in GCO does not occur exactly at \emph{T}\textsubscript{N}, rather it follows the magnetic phase transitions at 21\,K and occurs at 16\,K. 
Our DFT+$U$ calculations further support this argument as we do not notice any phonon instability when the magnetic order is changed from FM to AFM.
This suggest that no structural phase transition should occur exactly at \emph{T}\textsubscript{N}. 
However, below \emph{T}\textsubscript{N} the system could undergo a structural phase transition due to the relaxation of stress and forces on atoms within the AFM phase~\cite{KFR2012}.

\section{Conclusions}
\label{sec:conclusions}
Results from our \ss{combined} experimental and computational investigations of the IR and Raman-active modes of the normal spinel GeCo\textsubscript{2}O\textsubscript{4}
\ss{with the effective spin $S$=1/2 ground state} 
have been presented here with the following major conclusions: (i) The measured frequencies of the IR and Raman-active modes at room temperature are in good agreement with the results obtained from our DFT+$U$ calculations. (ii) All the IR and Raman-active modes exhibit moderate spin-phonon coupling in GeCo\textsubscript{2}O\textsubscript{4}. (iii) The temperature dependence of the Raman-active modes carried out between 5\,K and 100\,K with a special attention given to the region between \emph{T}\textsubscript{N} ($\sim$ 21\,K) and \emph{T}\textsubscript{S} ($\sim$ 16\,K) shows noticeable anomalies in the line parameters of the Raman-active modes. 
(iv) The temperature-dependent frequency shift of a shoulder appearing near the peak of the Raman-active mode $T$\textsubscript{2g}(1) validates that the structural phase transition in GeCo\textsubscript{2}O\textsubscript{4} is distinct from the magnetic phase transition occurring at \emph{T}\textsubscript{N}. 
%Our DFT+$U$ calculations further support this claim. 
Investigations of the temperature dependence of the IR modes covering the region below \emph{T}\textsubscript{N} is recommended since it is likely to provide significant information on the transitions at \emph{T}\textsubscript{N} and \emph{T}\textsubscript{S}.

Our DFT+$U$ calculations reveal that  exchange interactions up to at least the third neighbors are required to correctly describe the low-temperature antiferromagnetic ordering in GeCo\textsubscript{2}O\textsubscript{4}. 
\ss{We find that the nearest-neighbor magnetic exchange interaction has a ferromagnetic nature and it is a superexchange interaction mediated {\it via} an intermediate oxygen ion having a Co-O-Co bond angle of  $\theta = 90^{0}$. 
Instead, the second and third near-neighbor exchange interactions are antiferromagnetic in nature, and they involve more than one ion along the exchange interaction path corresponding to the super-super exchange interaction. 
These interactions play a vital role in stabilizing the ($\textbf{q}$= $\frac{1}{2},\frac{1}{2},\frac{1}{2}$) antiferromagnetic order in GeCo\textsubscript{2}O\textsubscript{4} at low temperatures.}

\ss{The presence of the spin $S$=1/2 ground state in GeCo\textsubscript{2}O\textsubscript{4} due to spin-orbit coupling and local Jahn-Teller distortion effects, discussed in detail in Ref.~\cite{pramanik2019magnetic}, gets additional support from the recently reported results in an Ising linear chain system CoNb\textsubscript{2}O\textsubscript{6} having a similar $S$ =1/2 ground state of  Co\textsuperscript{2+} ions~\cite{thota2021}. Lastly, we note that inclusion of the spin-orbit coupling and local Jahn-Teller distortion effects in DFT+$U$ calculations may slightly change the quantitative values reported in this work without affecting the overall physics of the studied system.}

\section*{ACKNOWLEDGEMENTS}
S.S., K.R., and D.V. acknowledge the support from Office of Naval Research (ONR) Grants N00014-16-1-2951 and N00014-19-1-2073. P.P. and S.G. acknowledge the FIST program of Department of Science and Technology, India for partial support of this work (Ref. No. SR/FST/PSII-020/2009 and SR/FST/PSII-037/2016).

\vspace{0.3 cm}

\textsuperscript{*}\,These authors contributed equally to this work.\\

\vspace{-0.2 cm}
Corresponding author(s): 

\textsuperscript{\textdagger}\,sobhit.singh@rutgers.edu,

\textsuperscript{\ddag} \,subhasht@iitg.ac.in

\bibliography{Ref}
\end{document}